\documentclass[%
 reprint,
superscriptaddress,
 amsmath,amssymb,
 aps,
prb,
]{revtex4-2}

\usepackage[normalem]{ulem}
\usepackage[dvipsnames]{xcolor}
\usepackage[hidelinks]{hyperref}
\usepackage{graphicx}

\usepackage{algorithm}
\usepackage[end]{algpseudocode}

\begin{document}

\title{Self-heating effects and switching dynamics in graphene multiterminal Josephson junctions}

\author{M\'at\'e Kedves}
\affiliation{Department of Physics, Institute of Physics, Budapest University of Technology and Economics, M\H uegyetem rkp.\ 3., H-1111 Budapest, Hungary}
\affiliation{MTA-BME Correlated van der Waals Structures Momentum Research Group, M\H uegyetem rkp.\ 3., H-1111 Budapest, Hungary}

\author{Tam\'as P\'apai}
\affiliation{Department of Physics, Institute of Physics, Budapest University of Technology and Economics, M\H uegyetem rkp.\ 3., H-1111 Budapest, Hungary}
\affiliation{MTA-BME Correlated van der Waals Structures Momentum Research Group, M\H uegyetem rkp.\ 3., H-1111 Budapest, Hungary}

\author{Gerg\H{o} F\"ul\"op}
\affiliation{Department of Physics, Institute of Physics, Budapest University of Technology and Economics, M\H uegyetem rkp.\ 3., H-1111 Budapest, Hungary}
\affiliation{MTA-BME Superconducting Nanoelectronics Momentum Research Group, M\H uegyetem rkp.\ 3., H-1111 Budapest, Hungary}

\author{Kenji Watanabe}
\affiliation{Research Center for Electronic and Optical Materials, National Institute for Materials Science, 1-1 Namiki, Tsukuba 305-0044, Japan}

\author{Takashi Taniguchi}
\affiliation{Research Center for Materials Nanoarchitectonics, National Institute for Materials Science,  1-1 Namiki, Tsukuba 305-0044, Japan}

\author{P\'eter Makk}
\email{makk.peter@ttk.bme.hu}
\affiliation{Department of Physics, Institute of Physics, Budapest University of Technology and Economics, M\H uegyetem rkp.\ 3., H-1111 Budapest, Hungary}
\affiliation{MTA-BME Correlated van der Waals Structures Momentum Research Group, M\H uegyetem rkp.\ 3., H-1111 Budapest, Hungary}

\author{Szabolcs Csonka}
\affiliation{Department of Physics, Institute of Physics, Budapest University of Technology and Economics, M\H uegyetem rkp.\ 3., H-1111 Budapest, Hungary}
\affiliation{MTA-BME Superconducting Nanoelectronics Momentum Research Group, M\H uegyetem rkp.\ 3., H-1111 Budapest, Hungary}
\affiliation{Institute of Technical Physics and Materials Science, Center for Energy Research,
Konkoly-Thege Mikl\'os \'ut 29-33., 1121, Budapest, Hungary}

\date{\today}

\begin{abstract}
We experimentally investigate the electronic transport properties of a three-terminal graphene Josephson junction. We find that self-heating effects strongly influence the behaviour of this multiterminal Josephson junction (MTJJ) system. We show that existing simulation methods based on resistively and capacitively shunted Josephson junction networks can be significantly improved by taking into account these heating effects. We also investigate the phase dynamics in our MTJJ by measuring its switching current distribution and find correlated switching events in different junctions. We show that the switching dynamics is governed by phase diffusion at low temperatures. Furthermore, we find that self-heating introduces additional damping which results in overdamped I-V characteristics when normal and supercurrents coexist in the device.
\end{abstract}

\maketitle

Multiterminal Josephson junctions (MTJJs) consisting of a single scattering region connected to multiple superconducting terminals attracted significant attention in recent years. Theoretical works showed that MTJJs may enable multiplet supercurrents\,\cite{Nowak2019,Melin2019,Melin2020,Melin2023,Doucot2020,Melo2022} and the Andreev bound state (ABS) spectra of MTJJs can exhibit non-trivial topology and simulate the band structure of Weyl semimetals\,\cite{Riwar2016,Eriksson2017,Houzet2019, Xie2017,Xie2018,Xie2022,Meyer2017,Deb2018,PeraltaGavensky2018,Klees2020,Fatemi2021,Peyruchat2021,Weisbrich2021,Chen2021,Chen2021a,Repin2022,PeraltaGavensky2023}. Although some of the theoretically proposed key features remain unobserved, recent experimental advances led to the observation of hybridized ABSs\,\cite{Coraiola2023, Matsuo2022, Matsuo2023, Matsuo2023b}, broken spin degeneracy and ground state parity transitions\,\cite{Coraiola2023a}, signatures of quartet supercurrents\,\cite{Pfeffer2014,Cohen2018,Arnault2022,Graziano2022,Huang2022}, the Josephson diode effect\,\cite{Chiles2023,Gupta2023, Matsuo2023a,Coraiola2024} and topological phase transitions\,\cite{Strambini2016}, highlighting the versatility of MTJJ devices. 

On the other hand, several experimental works found that the transport characteristics of MTJJs can be reasonably well modeled by a network of resistively and capacitively shunted Josephson junctions (RCSJ), in which each pair of terminals is connected by an RCSJ element. This relatively simple approach is able to qualitatively capture features of current-biased measurements, such as the coexistence of normal and supercurrents between different terminals\,\cite{Draelos2019,Arnault2022, Graziano2022,Koelzer2023,Zhang2023} and multiplet resonances\,\cite{Arnault2022,Zhang2023}. In spite of some agreement between simulations and measurements, these models in general fail to quantitatively capture the observations when normal and supercurrents coexist in the scattering region. This lack of agreement can be attributed to heating effects due to the presence of normal currents\,\cite{Draelos2019} that influence the supercurrent flowing in other parts of the device. Furthermore, the observation of more exotic phenomena, such as multiplet supercurrents\,\cite{Arnault2022, Zhang2023} and quantized transconductance\,\cite{Eriksson2017,Riwar2016}, also rely on the presence of finite voltages between some of the terminals that necessarily imply the existence of normal currents and heating effects. Due to the large superconducting gap $\Delta$ of the terminals which prevents the outflow of hot electrons, these heating effects can significantly modify the superconducting properties of MTJJs. Moreover, heating effects can have an impact on the switching dynamics of single Josephson junctions\,\cite{Clarke1988, Lee2011} which could be enhanced in the case of MTJJs, due to the complex geometry and the non-trivial current distribution.

In this work, we experimentally investigate a three-terminal graphene Josephson junction and compare our current-biased measurements to an RCSJ network model which enables us to identify the limitations of these models. Next, we present an improved simulation method, incorporating heating effects due to the presence of normal currents which results in a significantly better agreement with the measurements. Furthermore, we investigate the switching dynamics of our device and observe a non-trivial behaviour of the switching current distribution (SCD) at low temperatures that is governed by phase diffusion. We find that this behaviour is also modified by the heating effects due to normal currents. Finally, we investigate the charge carrier density dependence of the measured and simulated resistance maps which gives us further insight into the possible cooling mechanisms via which the dissipated heat escapes from the device.

\begin{figure*}
\includegraphics[width=2\columnwidth]{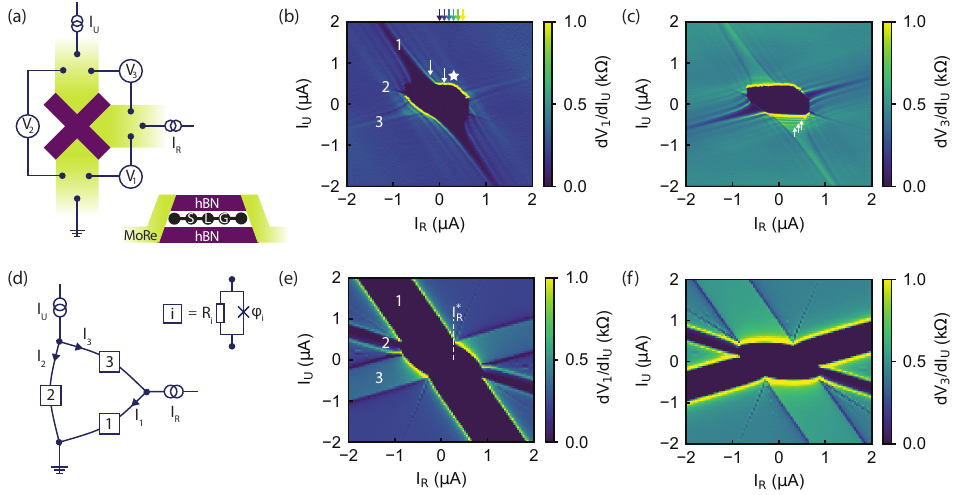}
\caption{\label{fig:1} a) Schematic representation of the multiterminal Josephson junction. Current biases $I_U$ and $I_R$ are applied via two separate contacts and the third contact is grounded. Voltages $V_i$ are measured between the three pairs of contacts. b,c) Differential resistance $dV_1/dI_U$ and $dV_3/dI_U$ as a function of the current biases. In panel b), white arrows illustrate the position of $T$-dependent measurement of $I-V$ curves (Fig.\,\ref{fig:3}.d) and colored arrows correspond to bias values where SCD measurements were performed (Fig.\,\ref{fig:3}.b). White star symbol shows the extended region where a finite voltage develops between all terminals simultaneously. d) RSJ network model of our device. White arrows in panel c) point to resonant features attributed to MAR. e,f) Simulated differential resistance maps analogous to panels b) and c), respectively. $I_R^{*}$ corresponds to the single current bias value of $I_R$ where all three junctions switch to normal state simultaneously as $I_U$ is ramped.}
\end{figure*}

Fig.\,\ref{fig:1}.a shows the schematic representation of our device. A cross-shaped hBN/graphene/hBN heterostructure is connected to three MoRe superconducting electrodes (optical image is shown in Appendix\,\ref{app:geom}). The separation of neighbouring contacts is around 150 nm. The charge carrier density $n$ in graphene can be tuned via the voltage applied to the doped Si substrate that acts as a global backgate, while a 300 nm thick SiO$_2$ layer forms the gate dielectric. In our experiments, one of the electrodes is grounded and two independent DC current biases, $I_R$ and $I_U$, are applied via the remaining two contacts, and differential voltages -- $V_{1}$, V$_{2}$ and V$_{3}$ -- between the three different pairs of terminals are measured. Transport measurements were carried out in a Leiden dilution refridgerator at a base temperature of 40 mK (unless otherwise stated). Fig.\,\ref{fig:1}.b(c) shows the differential resistance $dV_{1}/dI_U$ ($dV_{2}/dI_U$) -- obtained from the measured $V_{1}$ ($V_{2}$) voltage by numerical differentiation with respect to the current bias $I_U$ -- as a function of $I_U$ and $I_R$ at a backgate voltage of $V_{BG} = 10\,$V. Two main features can be identified in such a differential resistance map, similarly to previous experiments\,\cite{Draelos2019, Pankratova2020, Graziano2022,Arnault2021,Arnault2022,Koelzer2023,Gupta2023,Chiles2023}. First, in the center, around small current bias values an extended superconducting region of zero resistance can be observed. Second, superconducting arms (labeled by 1, 2 and 3 in Fig.\,\ref{fig:1}.b) are spreading out from this central superconducting region in multiple directions. Comparing differential resistance maps obtained from the measurements of $V_{1}$ (Fig.\,\ref{fig:1}.b), $V_{2}$ (Supplementary Information) and $V_{3}$ (Fig.\,\ref{fig:1}.c), it is easy to realise that the central superconducting region is present in all cases indicating that the whole sample is superconducting and supercurrent can flow between all of the terminals. On the other hand, each of the superconducting arms correspond to supercurrent flowing between only two terminals, resulting in zero resistance in only one of the differential resistance maps while a finite voltage develops between the remaining pairs of terminals (e.g. the SC arm labeled by 1 shows zero resistance in Fig.\,\ref{fig:1}.b and a finite voltage develops in Fig.\,\ref{fig:1}.c). This indicates that both normal and supercurrents can flow in the sample simultaneously. 

Previous works\,\cite{Draelos2019,Koelzer2023,Graziano2022,Arnault2021} showed that MTJJs can be described to a large extent by a network of RCSJ elements. Here, we neglect capacitive effects and model our three-terminal JJ with three resistively shunted junctions (RSJs), one between any pair of contacts, as shown in Fig.\,\ref{fig:1}.d. First, we present the results of this model and highlight its limitations in comparison with our measurements. Later, we show that the agreement between measurement and simulation can be improved by including self-heating effects in the model. As detailed in the Supplementary Information, the differential equations of this network model can be constructed from the Josephson equations and Kirchhoff's laws. The necessary input parameters of the model are the resistances ($R_i$ with $i\in \{1,2,3 \}$) and the critical currents ($I_{c,i}$) of the individual junctions. $R_i$ can be obtained from the measured differential resistances in the normal state, at large bias currents. For these, we obtain $R_1 = 420\,\Omega$, $R_2 = 1355\,\Omega$ and $R_3 = 815\,\Omega$. Furthermore, assuming that our junctions are in the short junction limit and using $I_{c,i}R_{i}\propto\Delta$, it is possible to extract $I_{c,i}$ from the measured differential resistance maps as well. For these, we get $I_{c,1}=545\,$nA, $I_{c,2}=170\,$nA and $I_{c,3}=280\,$nA, respectively. (See Appendix\,\ref{app:sim} for details on the extraction of parameters.) 

By numerically solving the set of differential equations for the network of Josephson junctions and resistors, we obtain differential resistance maps as shown in Figs.\,\ref{fig:1}.e and \ref{fig:1}.f. The model is capable of capturing the most prominent features of the measured differential resistance map: (i) the central superconducting region and (ii) the superconducting arms, corresponding to the coexistence of normal current and supercurrent. In the context of this model, the SC arms can be further discussed. The total current between any pair of terminals ($I_1$, $I_2$ and $I_3$) is determined by the Kirchhoff and Josephson equations for a given $I_U$ and $I_R$. It can be shown that for arbitrary $I_R$, a single value of $I_U$ exists for each junction for which the total junction current $I_i=0$ (see Appendix\,\ref{app:sim}). The ratio of $I_U/I_R$ for which $I_i=0$ is determined solely by the normal resistances and is independent of $I_U$ and $I_R$. Therefore, we expect to observe superconductivity in the vicinity of lines with slopes defined by the normal resistances. We also note that, in this particular geometry, due to Kirchhoff's law which states that the sum of voltages in a closed loop has to be zero, a single junction cannot switch to the normal state alone, a voltage drop has to appear on either two or all three junctions simultaneously. Therefore, outside the central SC region, the SC arms correspond to a configuration where only a single junction is superconducting and the other two resides in the normal state. On the other hand, several missing features can also be identified in the simulated resistance maps. The most prominent example is the decay of superconductivity that can be observed in the measurements along the superconducting arms. While the width of these arms in the simulated maps is constant towards higher current bias values, in the measurements a clear narrowing of the zero-resistance regions can be observed. Furthermore, in the measured resistance maps an extended region exists where all three junctions switch to the normal state simultaneously (e.g. marked by star symbol in Fig.\,\ref{fig:1}.b), whereas in the simulated maps, this simultaneous switching of all three junctions can only be observed for a single bias value $I_{R}^{*}$ (marked also by vertical dashed line in Fig.\,\ref{fig:1}.e). Finally, multiple resonant features (e.g. marked by white arrows in Fig.\,\ref{fig:1}.c) are visible in the measurements parallel to the superconducting arms that are attributed to multiple Andreev reflections (MAR)\,\cite{Pankratova2020} which cannot be accounted for by our simple model. We also note that MAR features are more pronounced for junction 1 than the other two junctions. This implies a larger contact transparency and may explain the larger critical current and lower resistance extracted for this junction.

\begin{figure}
\includegraphics[width=1\columnwidth]{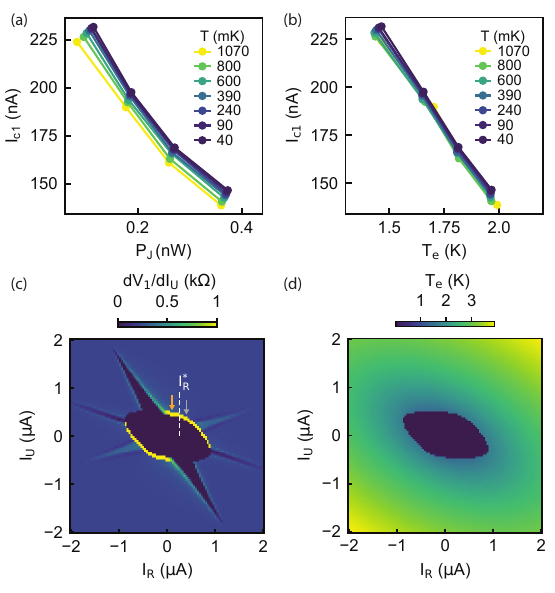}
\caption{\label{fig:2} Critical current of junction 1 $I_{c,1}$ along the corresponding superconducting arm as a function of a) heating power $P_J$ and b) electronic temperature $T_e$ calculated assuming only phonon cooling. c) Simulated differential resistance map taking into account the elevated electronic temperature due to normal current flowing in the device. d) Simulated map of $T_e$ as a function of current biases.}
\end{figure}

The narrowing of the superconducting arms is attributed to Joule heating from the dissipative normal currents in the scattering region\,\cite{Draelos2019}. Due to the large superconducting gap of the MoRe that prevents hot electron diffusion towards the leads, the electron system can only dissipate heat via electron-phonon coupling. In this case, the dissipated power towards the substrate is given by $P_{e-ph}=\Sigma(T_{e}^{\delta}-T^{\delta})$\,\cite{McKitterick2016}, where $\Sigma$ is the electron-phonon coupling constant, $T_e$ and $T$ are the electron and phonon bath temperatures, respectively. Following along the lines of Ref.\,\cite{Draelos2019}, we determine $\Sigma$ from the temperature dependence of $I_{c,1}$ along the corresponding SC arm. For this, we measure the switching current $I_{s,1}$ for junction 1 by sweeping $I_U$ at different values of $I_R$ and bath temperatures. $I_{s,1}$ is then defined as the value of $I_U$ where $V_1$ crosses a certain threshold voltage (20 $\mu$V) corresponding to the switching from the SC to the normal state. As mentioned earlier, in this current-biasing scheme, $I_U$ and $I_R$ do not directly correspond to the junction currents $I_1$, $I_2$ or $I_3$. However, since along the SC arm supercurrent only flows in junction 1, it is possible to calculate the junction's critical current $I_{c,1}$ from $I_{s,1}$ (see Appendix\,\ref{app:sim}). Moreover, as it is detailed later, the switching current of a Josephson junction is prone to fluctuations due to thermal effects. To eliminate these fluctuations, we take the average of 10\,000 measurements to determine the average switching current $\overline{I}_{s,1}$. Next, we calculate the power $P_J$ dissipated in the normal regions from Joule heating as $P_J = I_U V_2$. Fig.\,\ref{fig:2}.a shows the measured critical current $I_{c,1}$ as a function of $P_J$ for different $T$ bath temperatures. As it can be seen from the figure, the increased heating power leads to the decrease of the switching current, similarly to the increased bath temperature. Our assumption is that $T_e$ is homogeneous in the device and the critical current value is defined by $T_e$ independently from whether it originates from bath heating or current dissipation. We then determine the value of $\Sigma$ for which $I_{c,1}$ as a function of the calculated equilibrium electron temperature $T_e$ scales onto a single curve. As it is discussed later, by assuming $\delta=4$, we obtain $\Sigma = 25\,$pW/K$^4$. This is shown in Fig.\,\ref{fig:2}.b, where all the curves fall on top of each other. 
Although it is challenging to determine the exact active area of our device, we estimate that $\Sigma$ scaled by the graphene’s area yields $\sim100\,$W/m$^2$K$^4$. This is an order of magnitude larger than the value obtained in reference\,\cite{Draelos2019} ($\sim10\,$W/m$^2$K$^3$) and significantly larger than the value obtained for large area, nonencapsulated graphene devices\,\cite{McKitterick2016} ($<50\,$mW/m$^2$K$^4$). The authors of reference\,\cite{Draelos2019} also speculate that electron-phonon coupling can be enhanced by the presence of the hBN substrate and by scattering at the edges of the graphene layer. Since our device area is about an order of magnitude smaller than the device studied in reference\,\cite{Draelos2019}, scattering at the edges could be even more significant and could explain the larger value obtained for $\Sigma$ scaled by the graphene’s area.

To take the effects of self-heating into account in our simulations, we perform a fixed-point iteration based on the RSJ model introduced previously. First, we perform the previous simulation with the experimentally determined $R_{i}$ and $I_{c,i}$ parameters for all $I_U$ and $I_R$. We then calculate the Joule heat dissipated in the whole network as $P_J=\sum_{i} V_{i}^{2}/R_i$. From $P_J$, we can obtain the equilibrium electron temperature $T_e$ using the electron-phonon coupling model for all $I_U$ and $I_R$ bias currents. Finally, we take into account the elevated temperature using an $I_{c}(T_e)$ function which we reconstruct from the temperature dependent measurements shown previously in Fig.\,\ref{fig:2}.a and \ref{fig:2}.b and from the temperature dependence of the central superconducting region (see Appendix\,\ref{app:sim} for more details). We then iterate this process to achieve a self-consistent solution, using the modified $I_{c,i}$ values in our RSJ model which now also depend on the applied $I_U$ and $I_R$ current biases. 

Fig.\,\ref{fig:2}.c shows the simulated $dV_1/dI_U$ map obtained in our model with self-heating. Compared to Fig.\,\ref{fig:1}.d, several improvements can be observed. First of all, the narrowing of the SC arms is qualitatively reproduced. The remaining quantitative difference could be explained by the incorrect reconstruction of the $I_c(T_e)$ function. Secondly, the improved simulation method is capable of producing an extended edge on the contour of the central SC region where all three junctions switch to the normal state simultaneously. It is also worth noting that the simulated resistance map is inversion symmetric in contrast to the measurements where the sweep direction of the bias currents results in a slightly asymmetric central SC region.  Finally, Fig.\,\ref{fig:2}.d shows the map of $T_e$, illustrating that the heating outside the central SC region is significant, increasing the equilibrium temperature to a few Kelvins, an order of magnitude above the bath temperature, in agreement with our measurements shown in Fig.\,\ref{fig:2}.b.

\begin{figure}
\includegraphics[width=1\columnwidth]{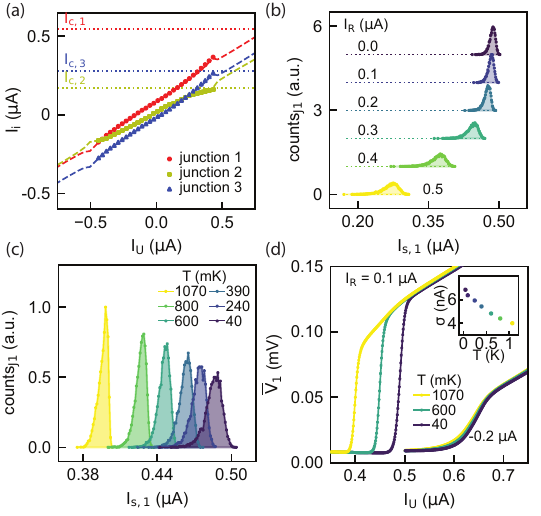}
\caption{\label{fig:3}  a) Current distribution between the three junctions in the central superconducting region obtained from the RSJ model (dashed lines) and from numerically minimising the Josephson energy of the whole network (symbols) for $I_R = 0.1\,\mu$A. Dotted lines show the critical current of each junction. b) SCD for junction 1 measured at different $I_R$ in the central superconducting region obtained from 10\,000 measurements. c) Temperature dependence of the SCD at $I_R = 0.1\,\mu$A. The narrowing of the SCD with increasing temperature is consistent with phase diffusion. d) Averaged I-V curves obtained from 10\,000 individual measurements in the central SC region ($I_R = 0.1\,\mu$A) and in the SC arm of junction~1 ($I_R = -0.2\,\mu$A) for different temperatures. Inset: standard deviation $\sigma$ of the SCD as a function of $T$ for $I_R = 0.1\,\mu$A.}
\end{figure}

In the following, we further investigate the interplay between the three junctions in the regions where all junctions switch to the normal state simultaneously around $I_{R}^{*}$ (also shown in Fig.\,\ref{fig:2}.c). Fig.\,\ref{fig:3}.a shows the current in each junction as a function of $I_U$ for $I_R=0.1\,\mu$A$\lesssim I_{R}^{*}$ (orange arrow in Fig.\,\ref{fig:2}.c) obtained from our simulation with self-heating (dashed lines). Although, $I_i$ cannot be obtained from our measurements, we can also calculate the current in each junction as long as all junctions are superconducting by numerically minimising the Josephson energy (symbols). As it is visible in Fig.\,\ref{fig:3}.a, this method is consistent with our simulation. The dotted horizontal lines show the critical current of the respective junctions. As it is visible also in Fig.\,\ref{fig:1}.e, junction 1 is far below its critical current when junction 2 and 3 reach their respective critical currents. Therefore, without taking self-heating into account, we only expect junction 2 and 3 to switch together. However, when heating is included (Fig.\,\ref{fig:2}.c and measurements on Fig.\,\ref{fig:1}.b and \ref{fig:1}.c), all three junctions switch at the same $I_U$. From these, we can infer that junction 2 and 3 switch together and junction 1 switches immediately afterwards due to heating from the other junctions. On the other hand, for $I_R\sim I_{R}^{*}$, all three junctions reach their critical currents simultaneously and heating should play no role in the switching process, while for $I_R\gtrsim I_{R}^{*}$ (gray arrow in Fig\,\ref{fig:2}.c), junction 1 and 2 switch together and junction 3 switches due to heating. Based on the previous arguments, we emphasize that the observation of this correlated switching of all three junctions in an extended region along the border of the central SC region is strong evidence for self-heating effects.

To gain insight into the dynamics of these correlated switchings, it is essential to investigate not only the average switching current but also its distribution. Fig.\,\ref{fig:3}.b shows the switching current distribution (SCD) obtained from the measurement of $V_1$ for different $I_R$ at 40 mK base temperature. The investigated values of $I_R$ are also indicated on top of Fig.\,\ref{fig:1}.b by coloured arrows. Each distribution is obtained by sweeping $I_U$ and detecting the switching current using the previously defined threshold voltage. This process is repeated 10\,000 times and a distribution of switching current values is obtained. Interestingly, we observe that the width of the SCD is greatly tunable by $I_R$ (Fig.\,\ref{fig:3}.b). We find that the standard deviation $\sigma$ of the SCD, which describes the width of the distribution, increases by a factor of 2. This broadening of the SCD could be explained by the different junctions that switch simultaneously at different $I_R$. As junction 1 takes over the role of junction 3 with increasing $I_R$, the sum of the critical currents of the two junctions that switch simultaneously increases which could lead to a wider distribution. It is also important to note that during the measurement of the SCD of junction 1, we simultaneously recorded the SCD of junction 2 obtained from the appearance of a finite $V_2$ and find that the distributions are identical and the switching events of the two junctions are indistinguishable within the time-scales of our measurement (see Appendix\,\ref{app:scd}). This suggests that the thermalisation of the device is faster than our data acquisition.

To further investigate the escape dynamics of our device, we measure the temperature dependence of the SCD along the contour of the central SC region. This is shown in Fig.\,\ref{fig:3}.c for $I_R = 0.1\,\mu$A and a similar trend is observed for all investigated values of $I_R$ inside the central SC region. It is clearly visible that the SCD gets narrower with increasing $T$ which is in stark contrast to the thermally activated behaviour as the SCD is expected to broaden with temperature. This is further confirmed by calculating $\sigma$ as a function of bath temperature for $I_R=0.1\,\mu$A that is shown in the inset of Fig.\,\ref{fig:3}.d. Here, a $\sim40$\% decrease of $\sigma$ is visible in the investigated temperature range. The observed narrowing of the SCD with increasing $T$ is a consequence of phase diffusion due to thermally activated escape and retrapping and is consistent with previous observations in moderately damped Josephson junctions\,\cite{Fenton2008} and planar Josephson junctions\,\cite{Haxell2023}. However, it is important to note that we observe the narrowing of the SCD in the whole available temperature range and do not find the broadening of the SCD due to thermally activated escape even for the lowest temperatures. This suggests that phase diffusion is significant even at base temperature.

We performed similar measurements along the SC arm of junction 1. Here, we find a different behaviour and we cannot resolve a clear SCD. Fig.\,\ref{fig:3}.d shows the averaged I-V curves of the 10\,000 individual measurements for $I_R=-0.2\,\mu$A and $I_R=0.1\,\mu$A (white arrows in Fig.\,\ref{fig:1}.b) for different temperatures. In the central SC region, for $I_R=0.1\,\mu$A, a sharp transition between the SC and normal states can be seen. In this case, the curvature of the averaged I-V curves results from averaging curves with fluctuating switching currents. On the other hand, for $I_R=-0.2\,\mu$A, along the SC arm of junction~1, a smooth transition is observed indicating that a finite voltage develops below the switching current. This is also consistent with the theoretical expectations for moderately damped Josephson junctions at higher temperatures\,\cite{Fenton2008}. As $T$ increases, the thermally activated retrapping results in a significant damping and the junctions become overdamped. This is further confirmed by the $T$-dependence of the curves. For $I_R=0.1\,\mu$A, as $T$ is increased, the switching current decreases (also visible in Fig.\,\ref{fig:3}.c). However, for $I_R=-0.2\,\mu$A, along the SC arm of junction 1, the effect of increasing $T$ is negligible, the increase of $T$ rather makes the transition between the SC and normal states smoother, as it is expected for overdamped junctions. It is also consistent with the self-heating picture, since increasing the bath temperature has less effect on the electronic temperature when a large heating power is already present due to the normal currents in the device. Therefore, we conclude that the switching of our multiterminal device is determined by phase diffusion at lower temperatures along the contour of the central SC region and show overdamped characteristics along the SC arm of junction 1 due to the increased temperature.

\begin{figure}
\includegraphics[width=1\columnwidth]{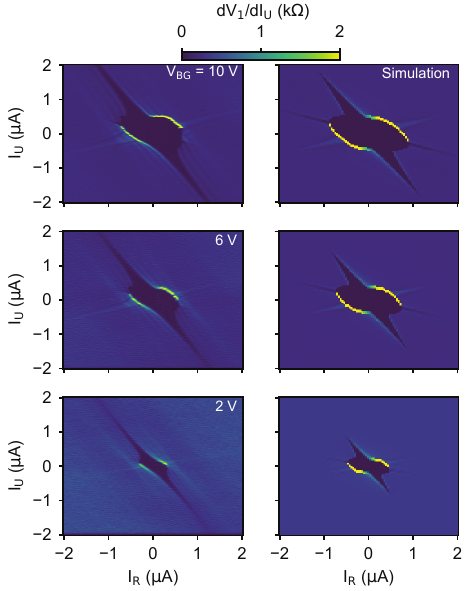}
\caption{\label{fig:4} Backgate dependence of the measured (left) and simulated (right) differential resistance maps. Simulations were performed with $\delta=4$ and $\Sigma=25\,$pW/K$^{4}$ for $V_{BG}=10\,$V and $\Sigma$ was scaled according to the $\sqrt{n}$-dependence expected for the clean limit of electron-phonon coupling.}
\end{figure}

Finally, we investigate the dependence of the differential resistance maps on the applied backgate voltage $V_{BG}$. As mentioned earlier, the exponent of the electron-phonon cooling power formula $\delta$ can be 3 or 4, depending on electronic mean free path $l_{\mathrm{mfp}}$ and temperature\,\cite{McKitterick2016}. At the relatively low temperatures accessed in our measurements, $\delta=4$ describes phonon cooling in clean devices where $l_{\mathrm{mfp}}$ is large, while $\delta=3$ corresponds to phonon cooling modified by impurity scattering in devices with small $l_{\mathrm{mfp}}$. Furthermore, the expression for $\Sigma$ is also different in the two limits. In the clean limit $\Sigma = \pi^2 D^2 |E_F| k_B^4/15 \rho_M \hbar^5 v_F^3 s^3$, where $D$ is the deformation potential of graphene which describes the electron-phonon coupling strength, $\rho_M$ is the mass density of graphene, $v_F = 10^6\,$m/s is the Fermi velocity, $E_F=\hbar v_F \sqrt{\pi n}$ is the Fermi energy and $s=2\times10^4\,$m/s is the speed of sound in graphene. It can easily be shown that in the clean limit $\Sigma\propto \sqrt{n}$, while in the dirty limit the expression is modified and $\Sigma$ becomes independent of $n$\,\cite{McKitterick2016}. Fig.\,\ref{fig:4} shows the measured and simulated resistance maps for different $V_{BG}$ and $\delta = 4$. We scale $\Sigma$ according to the $\sqrt{n}$-dependence and $n$ is calculated according to a planar capacitor model based on the hBN and SiO$_2$ dielectrics (see Appendix\,\ref{app:add_sim}). From Fig.\,\ref{fig:4}, it is visible that the qualitative trend is reproduced well. Here, we assumed $\delta=4$, but note that a reasonably good agreement can also be achieved by taking $\delta=3$ and a constant $\Sigma=30\,$pW/K$^{3}$ (see Appendix \ref{app:add_sim}). Although the overall qualitative agreement between measurement and simulation is good, some differences can still be observed. Most notably, some of the SC arms persist up to larger current bias values in the measurements, especially noticable for $V_{BG}=2\,$V. This could be explained by the appearance of additional cooling paths. As $T_e$ is increased up to a few Kelvins, $k_B T_e$ becomes comparable to $\Delta$, allowing quasiparticles to diffuse into the MoRe leads. Furthermore, we assumed that $T_e$ is homogeneous in the whole device which does not necessarily hold for large heating powers. The inhomogeneity of $T_e$ could significantly modify the ratio of normal and SC segments of the scattering region and, as a result, the estimated input parameters of our model would become increasingly inaccurate with increasing heating powers.

In conclusion, we have measured three-terminal graphene Josephson junctions and investigated the heating effects and junction dynamics in this multiterminal system. We have shown that a significant improvement can be achieved over existing RCSJ models for MTJJs by incorporating heating effects into the simulation method. By considering only Joule heating from the normal currents in the device and electron-phonon coupling as cooling mechanism, we were able to obtain the narrowing of the SC arms that is commonly observed in experiments and the simultaneous switching of all junctions. By measuring the charge carrier density dependence of the differential resistance maps, we could infer the limitations of our model, and suggest that, for significantly increased electronic temperatures, new cooling mechanisms might become available. We propose that by including additional cooling terms, such as the outflow of hot electrons via the SC terminals, our model could be further improved. Furthermore, from the investigation of the SCD, we concluded that the switching from the central SC region to the normal state is governed by phase diffusion even at very low temperatures. As the temperature is increased due to self-heating, this phase diffusion modifies the characteristics of the device, resulting in smooth I-V curves resembling overdamped Josephson junctions. Building on these results, future experiments could focus on the phase-biasing of MTJJs and inductance measurements using RF techniques in the SC state, where self-heating effects are absent. 

\section*{Author contributions}
M.K. and T.P. fabricated the device. Measurements were performed by M.K. and T.P. with the help of P.M. and Sz.Cs. M.K. and T.P. did the data analysis. Simulations were performed by G.F., M.K. and T.P. M.K. and P.M. wrote the paper and all authors discussed the results and worked on the manuscript. K.W. and T.T. grew the hBN crystals. The project was guided by P.M. and Sz.Cs.

\begin{acknowledgments}
This work acknowledges support from the Topograph, MultiSpin and 2DSOTECH FlagERA networks, the OTKA K138433 and K134437 grants, the VEKOP 2.3.3-15-2017-00015 grant and the EIC Pathfinder Challenge grant QuKiT. This research was supported by the Ministry of Culture and Innovation and the National Research, Development and Innovation Office within the Quantum Information National Laboratory of Hungary (Grant No. 2022-2.1.1-NL-2022-00004), by the FET Open AndQC and SuperGate networks and by the European Research Council ERC project Twistrain. We acknowledge COST Action CA 21144 superQUMAP. 
K.W. and T.T. acknowledge support from the JSPS KAKENHI (Grant Numbers 20H00354 and 23H02052) and World Premier International Research Center Initiative (WPI), MEXT, Japan.
\end{acknowledgments}

\appendix
\renewcommand{\theequation}{A\arabic{equation}}
\renewcommand{\thetable}{A\arabic{table}}

\section{\label{sec:device}Device geometry and measurement setup}
\label{app:geom}
\begin{figure}[ht]
\includegraphics[width=\columnwidth]{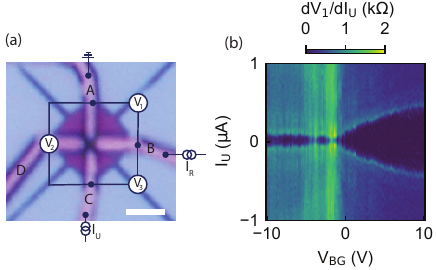}
\caption{\label{fig:S1} a) Optical microscopic image of the device with the schematic illustration of the measurement geometry. Scale bar is 2 $\mu$m. b) Differential resistance of junction 1 $dV_{1}/dI_{U}$ as a function of current bias $I_U$ and backgate voltage $V_{BG}$ for $I_R = 0$.}
\end{figure}

The measured sample is shown in Fig.\,\ref{fig:S1}.a. The dry-transfer technique with PC/PDMS
stamps was employed to stack hBN (20 nm, top)/SLG/hBN
(35 nm, bottom). To fabricate electrical contacts, we used electron beam lithography patterning followed by a reactive ion etching step using CHF$_3$/O$_2$ mixture and finally deposited MoRe (50 nm) by dc  sputtering. As it is visible on the optical microscopic image in Fig.\,\ref{fig:S1}.a, four MoRe contacts were fabricated, however, one of the contacts failed to contact the graphene layer, resulting in a three terminal device as it is presented in the main text. The separation of neighbouring contacts is around 150 nm. The heterostructure around the cross-shaped region was etched away using reactive ion etching with SF$_6$/O$_2$ mixture.

\begin{figure*}
\includegraphics[width=2\columnwidth]{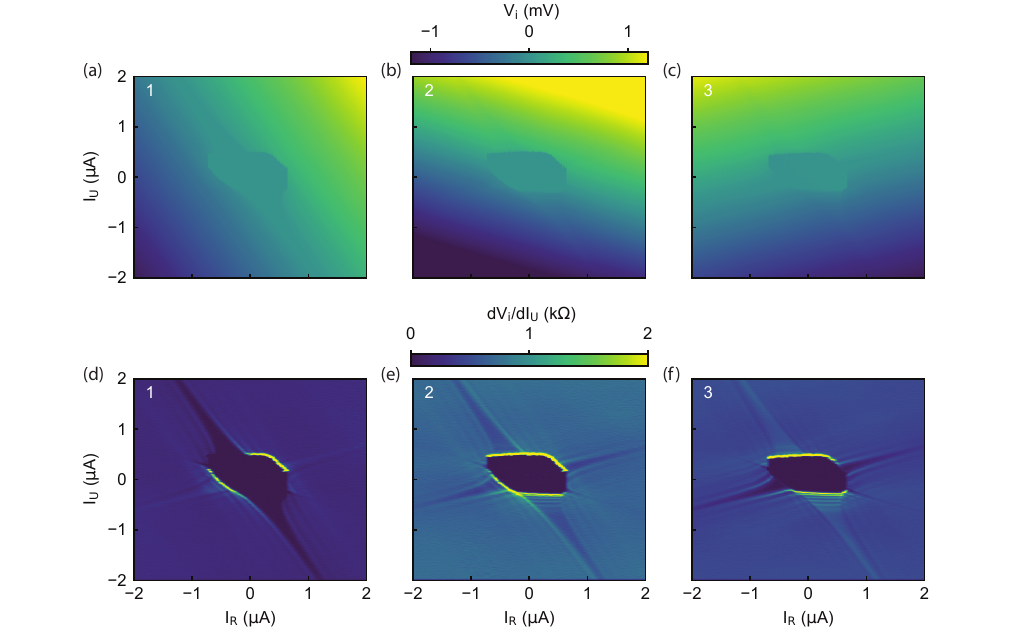}
\caption{\label{fig:S2} a-c) Raw measured voltages $V_{1}$, $V_{2}$ and $V_{3}$ as a function of $I_U$ and $I_R$. e-f) Differential resistances calculated from panels a-c)}
\end{figure*}

Transport measurements were carried out in a Leiden dilution refridgerator at a base temperature of 40 mK (unless otherwise stated). Measurements were performed using a NI USB 6341 measurement card. In each measurement, contact $A$ was grounded and the DC current biases $I_R$ and $I_U$ were applied via 1 M$\Omega$ preresistors to contact B and C, respectively. During the SCD measurements for fixed values of $I_R$, $I_U$ was ramped from 0 to 1\,$\mu$A at ramp rate of 100 $\mu$A/s while the voltages between two different pairs of terminals were simultaneously measured. Fig.\,\ref{fig:S2} shows the measured raw voltages as a function of the current biases, corresponding to the differential resistance maps shown in Fig.\,\ref{fig:1} of the main text.

Fig.\,\ref{fig:S1}.b shows the differential resitance of junction 1 as a function of $V_{BG}$ and $I_{U}$ showing a highly tunable critical current with $V_{BG}$ as it is common for graphene devices. The critical current can be tuned to zero near the charge neutrality point and we observe a significantly smaller critical current for negative $V_{BG}$ which we attribute to doping from the MoRe contacts and formation of a p-n junction at the MoRe interface\,\cite{Calado2015}.

\section{\label{sec:RSJ}RSJ simulations}
\label{app:sim}
As discussed in the main text, we start our simulation by solving an RSJ network model. For our three-terminal device, this consists of three blocks of resistively shunted Josephson junctions as shown in Fig.\,\ref{fig:1}.d of the main text. The $i-$th block is described by a resistor with resistance $R_i$ and the phase difference of the Josephson junction $\varphi_i$. The normal current flowing in the resistor is given by $I_{N,i} = V_i/R_i$, where $V_i$ is the voltage drop on the RSJ block. We employ a sinusoidal current-phase relation and the supercurrent flowing in the Josephson junction is given by $I_{s,i} = I_{c,i}\sin\varphi_i$. According to the corresponding Josephson equation, the time derivative of the phase difference is given by $\dot{\varphi_i}=2eV_i/\hbar$. With these, one can obtain the differential equation of a single RSJ block:
\begin{equation*}
    I_i = I_{c,i}\sin\varphi_i+\frac{\hbar}{2eR_i}V_i,
\end{equation*}
where $I_i$ is the total current flowing in the $i-$th block. Introducing the external current biases $I_U$ and $I_R$ and the superconducting phases of the corresponding leads $\varphi_U$ and $\varphi_R$ according to Fig.\,\ref{fig:S4}.a, choosing the phase of the grounded terminal as zero and applying Kirchhoff's law, one can end up with a set of coupled differential equations for the complete RSJ network:
\begin{equation} \label{eq1}
\begin{split}
\frac{da_1}{dt} & = \frac{2e}{\hbar}\left[I_U-I_{c,2}\sin\left(-\varphi_U\right) -I_{c,3}\sin\left(\varphi_R-\varphi_U \right)\right], \\
\frac{da_2}{dt} & = \frac{2e}{\hbar}\left[I_R-I_{c,1}\sin\left(-\varphi_R\right) +I_{c,3}\sin\left(\varphi_R-\varphi_U \right)\right],
\end{split}
\end{equation}

where

\begin{equation*}
\begin{split}
    a_1 = \frac{\varphi_R-\varphi_U}{R_3}-\frac{\varphi_U}{R_2},\\
    a_2 = -\frac{\varphi_R-\varphi_U}{R_3}-\frac{\varphi_U}{R_1}
\end{split}
\end{equation*}
and we made use of the fact that $\varphi_3=\varphi_R-\varphi_U$. By numerically solving equation system\,\ref{eq1}, we obtain the stationary $\varphi_i$ phase differences and $V_i$ voltages from which both the normal $I_{n,i}$ and supercurrents $I_{s,i}$ in each block can be calculated for a given $I_U$ and $I_R$.

\subsection{\label{subsec:params}Determination of junction parameters}
\begin{figure}
\includegraphics[width=\columnwidth]{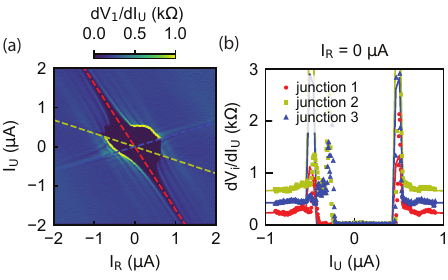}
\caption{\label{fig:S4} a) Differential resistance maps $dV_1/dI_U$. Dashed lines illustrate the obtained slopes of the SC arms. b) Measured differential resistances for $I_R=0$ (markers). Solid lines show the simulated differential resistances with our improved method, taking self-heating effects into account.}
\end{figure}

As mentioned in the main text, to quantitatively match the simulations to our measurement, we determine $R_i$ and $I_{c,i}$ from the measured differential resistance maps. First of all, it is easy to show that the ratio $I_U/I_R$ for which $I_{s,i}=0$, corresponding to the slope of the SC arms, is determined by the normal resistances as:

\begin{equation}\label{eq2}
\begin{split}
    \alpha & = -\frac{R_2+R_3}{R_2},\\
    \beta & = -\frac{R_1}{R_1+R_2},\\
    \gamma & = \frac{R_1}{R_2},
\end{split}
\end{equation}
for junctions 1, 2 and 3 respectively. For these, we obtain $\alpha=-1.6$, $\beta=-0.34$ and $\gamma=0.31$ from the measured differential resistance maps at $V_{BG}=10\,$V. These are shown with dashed lines in Fig.\,\ref{fig:S4}.b. Since these equations are not independent, we also calculate the differential resistances in the normal state where only normal currents are flowing as:
\begin{equation}\label{eq3}
\begin{split}
    R_I & = \frac{dV_1}{dI_U} = \frac{R_1R_2}{R_1+R_2+R_3},\\
    R_{II} & = \frac{dV_2}{dI_U} = \frac{R_2(R_1+R_3)}{R_1+R_2+R_3},\\
    R_{III} & = \frac{dV_2}{dI_U} = \frac{R_2R_3}{R_1+R_2+R_3}.
\end{split}
\end{equation}
Combining equation systems\,\ref{eq2} and \ref{eq3}, one can show that $R_2 = R_I(\gamma-\alpha)/\gamma$ and the normal resistances can be calculated. For these, we obtain $R_1 = 420\,\Omega$, $R_2 = 1355\,\Omega$ and $R_3 = 815\,\Omega$, respectively. Having obtained the normal resistances, it is also possible to calculate the junction critical currents $I_{c,i}$. First, we calculate the superconducting coherence length in graphene. Since the length of our junctions is smaller than 200 nm, well below the typical mean free path for similar graphene devices, we assume ballistic conduction. Using $\Delta=1.2\,$meV for the SC gap of the MoRe contacts\,\cite{Borzenets2016,Indolese2018}, the coherence length is given by $\xi=\hbar v_F/\pi\Delta\approx 200\,$nm. Therefore, we conclude that our junctions are in the short, ballistic limit which implies that $I_{c,i}R_{i}\propto\Delta$. This allows us to calculate $I_{c,i}$ from the measured differential resistance maps, using the previously calculated normal resistances. Using this, it can be shown that for $I_R=0$, the total critical current is given by $I_{c,tot}=I_{c,2}+I_{c,3}=I_{c,2}(1+R_2/R_3)$ and the individual junction critical currents $I_{c,i}$ can be calculated using the $R_{i}$ normal resistances. We associate $I_{c,i}$ with the values obtained from the differential resistance maps measured at base temperature. For these, we obtain $I_{c,1}=545\,$nA, $I_{c,2}=170\,$nA and $I_{c,3}=280\,$nA.
    
\begin{figure}
\includegraphics[width=\columnwidth]{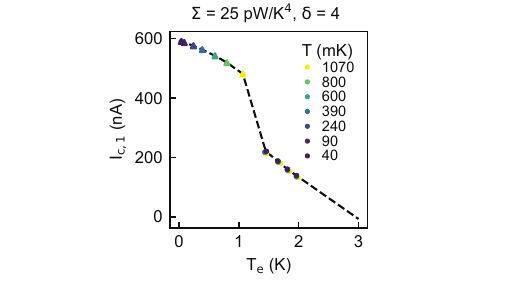}
\caption{\label{fig:S5} a) The experimentally obtained $I_{c,1}(T_e)$ function. Triangles show $I_{c,1}$ obtained from the central SC region and circles correspond to the values extracted from the SC arm of junction 1.}
\end{figure}

\begin{figure*}
\includegraphics[width=2\columnwidth]{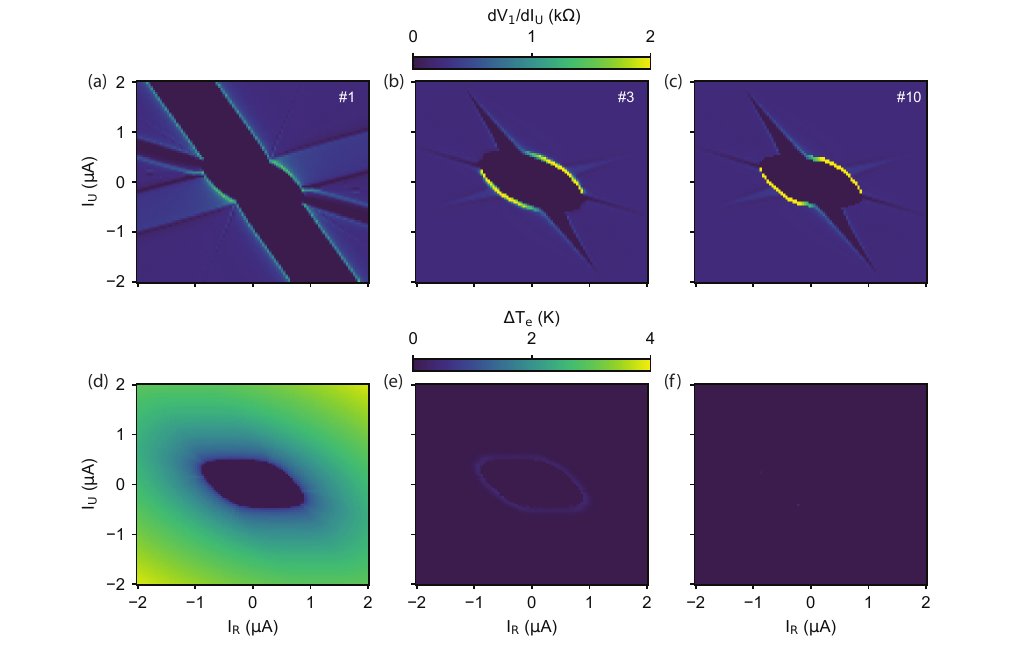}
\caption{\label{fig:S6} a-c) Simulated differential resistance maps after 1, 3 and 10 iterations, respectively. d-e) Change of electronic temperature $\Delta T_e = T_{n} - T_{n-1}$, where $n$ is the iteration step.}
\end{figure*}

\subsection{\label{subsec:IcT}Determination of $I_{c,i}(T_e)$}
As discussed in the main text, to include heating effects in our simulations, we perform a fixed-point iteration. The pseudocode for this algorithm is shown in Algorithm\,\ref{algorithm}. First, we solve the RSJ network model with the experimentally obtained parameters and calculate the Joule heating power as $P_J = \sum_{i}V_i^2/R_{i}$ and the equilibrium electron temperature as $T_e=\sqrt[4]{T^{4}+P_J/\Sigma}$, using $\Sigma=25\,$pW/K$^4$ as obtained from the temperature-dependent measurements (see Fig.\,\ref{fig:2}. of the main text) and assuming homogeneous temperature distribution in the  device. The next step is to take the effect of the elevated electron temperature into account via the $I_c(T_e)$ dependence. We construct this function from our temperature dependent measurements. For this, we have to consider two different regimes. First, in the central SC region, as discussed previously, the individual junction critical currents can be calculated using the normal resistances. Assuming that the ratio of the resistances does not change with temperature, we can obtain $I_{c,1}$ by taking $I_{c,tot}$ as the mean of the SCDs measured at $I_R=0$ for different $T$ (Fig.\,\ref{fig:3}.c. of the main text). Moreover, since in this region all junctions are superconducting, we can take $T_e = T$ as there is no Joule heating. 

Next, we consider the SC arm of junction 1. Utilizing the previous definition of the slope $\alpha$ of the SC arm of junction 1, for a given $I_R$ the supercurrent in junction 1 is zero for $I_U=\alpha I_R$. Furthermore, since along the SC arm, only junction 1 is superconducting and the remaining two junctions are in the normal state, we can calculate the ratio of $I_U$ that is flowing towards junction 1. Combining these, the net current of junction 1 is given by $I_1=(I_U-\alpha I_R) R_2 /(R_2+R_3)$. In this case, we define the average switching current of junction~1 $\overline{I}_{s,1}$ as the value of $I_U$ for which $\overline{V}_1$ exceeds the pre-defined threshold voltage (20\,$\mu$V), where $\overline{V}_1$ is the average voltage obtained from averaging 10000 individual measurements. From this, we calculate the critical current of junction 1 as $I_{c,1}=(\overline{I}_{s,1}-\alpha I_R)R_2/(R_2+R_3)$. The obtained values of $I_{c,1}$ for different $T_e$ are shown in Fig.\,\ref{fig:S5}. To find the value of $I_{c,1}$ for any $T_e$, we linearly interpolate and extrapolate. Finally, to get $I_{c,2}$ and $I_{c,3}$, we simply scale the $I_{c,1}(T_e)$ function according to the ratio of normal resistances, based on our previous arguments.

The simulated differential resistances for $I_R=0$ are shown with solid lines in Fig.\,\ref{fig:S4}.c. As it is visible, the simulated curves qualitatively match the measured points for $I_U>0$. For negative $I_U$, the retrapping to the SC state happens later in the measurements than in the simulations. We attribute this also to the elevated temperature due to self-heating, as the simulated curves do not take into account the sweep direction of the current bias.

\begin{figure*}
	\begin{minipage}{\linewidth}
		\begin{algorithm}[H]
			\caption{Iterative procedure for the self-consistent calculation of  junction currents and electronic temperature}\label{algorithm}
			\begin{algorithmic}
				\Function{calculate\_mtjj}{$n_{\rm{iter}} = 10$, $\delta$, $\Sigma$, $R_i$, $I_{c,i}(T_e)$ for $i \in \{1,2,3\}$, $T_{\rm{bath}}$}
				
				\State $T_e \leftarrow T_\mathrm{bath}$ for all $I_U$, $I_R$ \Comment{Initialization}
				
				\For{$n_{\rm{iter}}$ repetitions}
				\For{all $I_U$, $I_R$ in the range}
				
				\State $V_i, I_i \leftarrow $ solve ODE set using $I_{c,i} = I_{c,i}(T_e)$
				\Comment{Eq. system \ref{eq1}}
				\State  $P \leftarrow V_1^2/R_1 +  V_2^2/R_2 + V_3^2/R_3$
				\State $T_e^{\mathrm{new}} \leftarrow$ solve $P =\Sigma\cdot\left(T_e^\delta-T^\delta_\mathrm{bath}\right)$ for $T_e$\Comment{Assumes $P_{e-ph}=P_J$}
				
				\EndFor
				\EndFor
				
				\State \Return $I_i(I_U,I_R), V_i(I_U,I_R), T_e(I_U,I_R)$ 
				\EndFunction
			\end{algorithmic}
		\end{algorithm}
	\end{minipage}
\end{figure*}

\subsection{\label{subsec:iter}Iteration process}
To further illustrate the fixed-point iteration method, we show the simulated differential resistance map $dV_1/dI_U$ after different numbers of iteration in Fig.\,\ref{fig:S6}. The first step (Fig.\,\ref{fig:S6}.a) corresponds to the simulation without taking heating into account, also shown in Fig.\,\ref{fig:1}.e and \ref{fig:1}.f of the main text. After three iterations (Fig.\,\ref{fig:S6}.b), the main features of the measured resistance maps are well reproduced. Fig.\,\ref{fig:S6}.c shows the final result after 10 rounds of iteration which only shows minor differences compared to Fig.\,\ref{fig:S6}.b. Fig.\,\ref{fig:S6}.d-f shows the change of electronic temperature $\Delta T_e = T_{n} - T_{n-1}$, where $n$ is the iteration step and $T_{0}=40\,$mK is the base temperature. It can be seen that while the electronic temperature is drastically modified for the first step, later iterations only result in minor changes indicating the convergence of our simulations.

\section{Additional simulations}
\label{app:add_sim}
\begin{figure*}
\includegraphics[width=2\columnwidth]{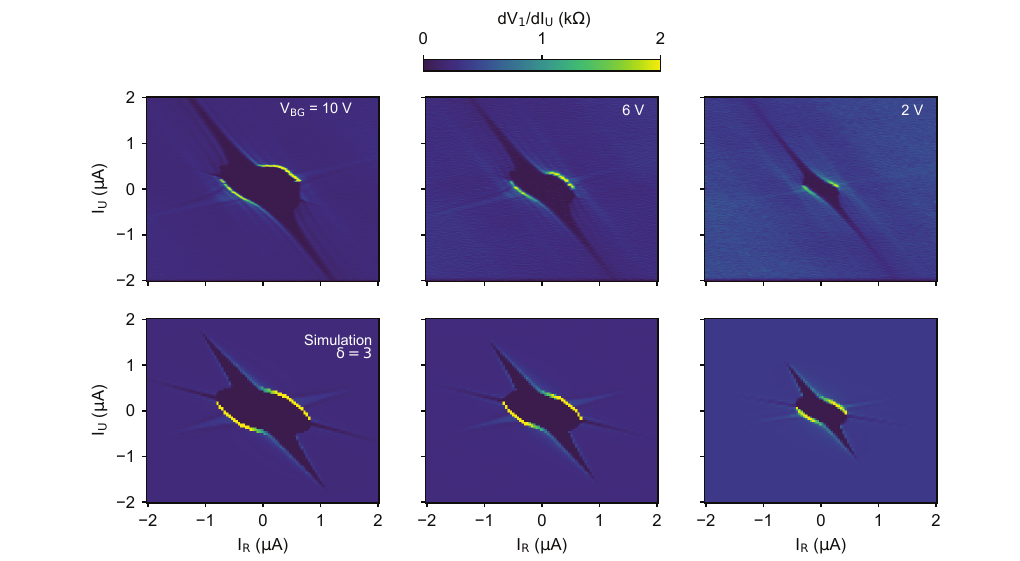}
\caption{\label{fig:S9} Simulation with $\delta=3$ and constant $\Sigma=30\,$pW/K$^3$.}
\end{figure*}

As mentioned in the main text, we can also perform the scaling of $I_{c,1}$ along the SC arm of junction 1 using $\delta=3$ corresponding to the dirty limit of electron-phonon coupling. This scaling yields $\Sigma=30\,$pW/K$^3$. We also construct the $I_c(T_e)$ function using this modified $\Sigma$ and simulate the differential resistance maps analogous to Fig.\,\ref{fig:4} of the main text. In this case, the expression for $\Sigma$ is modified, it is given by $\Sigma=\frac{2\zeta(3)D^2|E_F|k_B^3}{\pi^2\rho_M\hbar^4 v_F^3 s^2 l_{mfp}}$. It can be shown that, in this case, $\Sigma$ is independent of $n$. The simulated resistance maps for $\delta=3$ and $\Sigma=30\,$pW/K$^3$ are shown in Fig.\,\ref{fig:S9}. 

\begin{table}[h]
\centering
\begin{tabular}{c |c |c} 
 $V_{BG}$ (V) & $n$ ($10^{12}\,$cm$^{-2}$) & $\Sigma_{\delta=4}$ (pW/K$^4$)\\
 \hline
 10 &  0.74 & 25\\ 
 \hline
 6 &  0.48 & 20\\
 \hline
 2 &  0.22 & 14\\
\end{tabular}
\caption{\label{tab:n}Charge carrier densities $n$ and $\Sigma$ in case of $\delta=4$ corresponding to the values of $V_{BG}$ for which the differential resistance maps were measured and simulated.}
\end{table}

As detailed in the main text, for $\delta=4$, $\Sigma$ is scaled according to a $\sqrt{n}$-dependence. The $\Sigma$ values for each $V_{BG}$ can be found in Table\,\ref{tab:n}. We also present the charge carrier densities $n$ for the different $V_{BG}$ values where the differential resistance maps were measured and simulated in Table\,\ref{tab:n}. We determine the backgate voltage of the charge neutrality point $V_{CNP}=-1.4\,$V from the gate-dependent measurement shown in Fig.\,\ref{fig:S1}.c. Using this, the carrier density is given by $n=\alpha_{BG}(V_{BG}-V_{CNP})$. The lever arm of the backgate is calculated according to a planar capacitor model as $\alpha_{BG}=\varepsilon_0/e\cdot(d_{\mathrm{SiO2}}/\varepsilon_{\mathrm{SiO2}}+d_{\mathrm{hBN}}/\varepsilon_{\mathrm{hBN}})^{-1}$, where $\varepsilon_0$ is the vacuum permittivity, $e$ is the elementary charge, $\varepsilon_{\mathrm{SiO2}}=4$ ($\varepsilon_{\mathrm{hBN}}=3.3$) and $d_{\mathrm{SiO2}}=300\,$nm ($d_{\mathrm{hBN}}=35\,$nm) are the dielectric constant and thickness of SiO$_2$ (hBN), respectively.

\section{\label{sec:MAR}Multiple Andreev reflections}
\label{app:mar}
\begin{figure*}
\includegraphics[width=2\columnwidth]{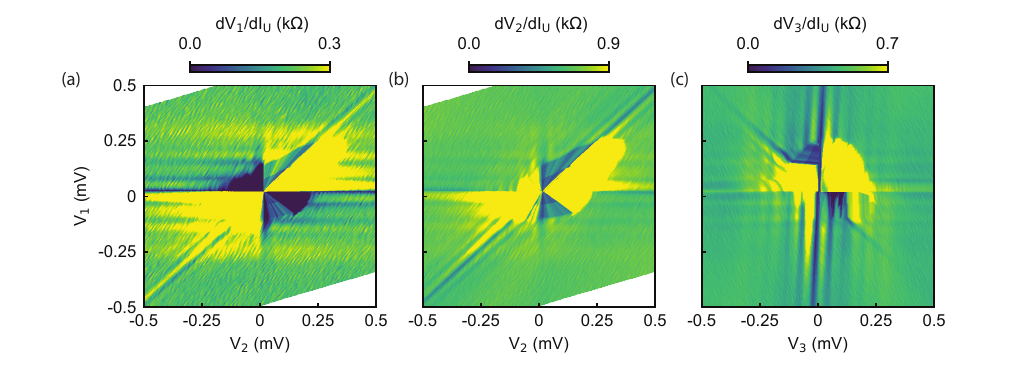}
\caption{\label{fig:S3} a) Measured differential resistances $dV_i/dI_U$ as a function of the measured voltages.}
\end{figure*}

Fig.\,\ref{fig:S3} shows the differential resitances $dV_{i}/dI_U$ plotted as a function of the measured voltages $V_i$. We observe resonant features that are attributed to multiple Andreev reflections\,\cite{Pankratova2020}. Each resistance map is plotted as a function of the two voltages that were measured simultaneously.

\section{Extended SCD data}
\label{app:scd}
\begin{figure}
\includegraphics[width=\columnwidth]{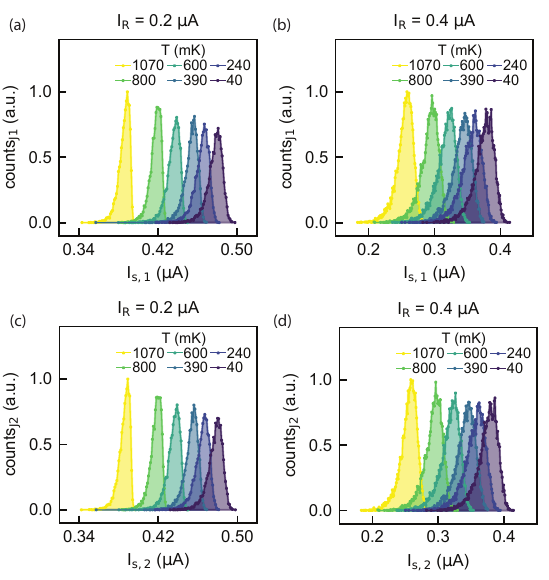}
\caption{\label{fig:S7} a-b) Additional SCD data for junction 1 measured at $I_R=0.2\,\mu$A and $I_R=0.4\,\mu$A, respectively, for different temperatures. c-d) SCDs for junction 2, simultaneously measured with the SCDs for junction 1.}
\end{figure}

As mentioned earlier, we performed the SCD measurements simultaneously for two different junctions. Fig.\,\ref{fig:S7}.a and \ref{fig:S7}.b shows additional SCDs for junction 1, while Fig.\,\ref{fig:S7}.c and \ref{fig:S7}.d shows the SCDs measured for junction 2. As mentioned in the main text, we observe similar tendencies for all investigated SCDs in the range of $0$\,$\mu$A$<I_R<0.5\,\mu$A. The narrowing of the SCDs with temperature can be observed for both junctions in the whole investigated temperature range. Furthermore, the SCDs obtained for junction 1 and 2 are almost identical, further showing that the two junctions switch in a correlated manner.

\section{Superconducting diode effect}
\label{app:diode}
\begin{figure}
\includegraphics[width=\columnwidth]{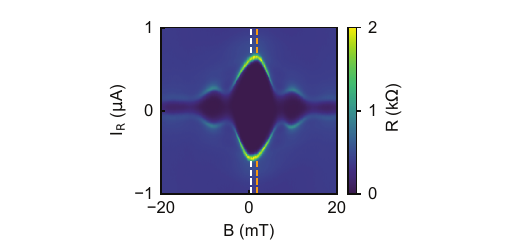}
\caption{\label{fig:S10}  Differential resistance $R$ as a function of out-of-plane magnetic field $B$ and $I_R$ for $I_U=0$. Orange and white dashed lines show the maximum of the switching and retrapping currents, respectively.}
\end{figure}

Previous works showed that MTJJs are a suitable platform to realise the Josephson diode effect\,\cite{Chiles2023,Coraiola2024,Gupta2023} where the amplitude of the critical current depends on the direction of the current flow. Fig.\,\ref{fig:S10} shows the differential resistance as a function of out-of-plane magnetic field $B$ and $I_R$ for $I_U=0$. The differential resistance is measured between contacts $A$ and $B$ using lock-in technique at 177 Hz frequency using 10 nA AC current bias applied via a 1 M$\Omega$ preresistor. During the measurement, $I_R$ is ramped from $-1\,\mu$A to $1\,\mu$A for fixed $B$. As it is visible in Fig.\,\ref{fig:S10}., the maximum switching and retrapping currents are observed for different $B$ (orange and white dashed lines, respectively) which is a signature of the Josephson diode effect.

\end{document}